\documentclass[sigconf]{acmart}
\AtBeginDocument{%
  }

\setcopyright{acmlicensed}
\copyrightyear{2018}
\acmYear{2018}
\acmDOI{XXXXXXX.XXXXXXX}
\acmConference[Conference acronym 'XX]{Make sure to enter the correct
  conference title from your rights confirmation email}{June 03--05,
  2018}{Woodstock, NY}
\acmISBN{978-1-4503-XXXX-X/2018/06}

\usepackage[utf8]{inputenc} 
\usepackage[T1]{fontenc}    
\usepackage{textgreek}      
\begin {document}

\title{AnomalyGen: An Automated Semantic Log Sequence Generation Framework with LLM for Anomaly Detection}

\author{Xinyu Li}
\email{lixy769@mail2.sysu.edu.cn}
\affiliation{%
  \institution{Sun Yat-sen University}
  \city{ZhuHai}
  \country{China}
}

\author{Yingtong Huo}
\email{ythuo@smu.edu.sg}
\affiliation{%
  \institution{Singapore Management University}
  \country{Singapore}
}

\author{Chenxi Mao}
\email{maochx5@mail2.sysu.edu.cn}
\affiliation{%
  \institution{Sun Yat-sen University}
  \city{ZhuHai}
  \country{China}
}

\author{Shiwen Shan}
\email{shanshw@mail2.sysu.edu.cn}
\affiliation{%
  \institution{Sun Yat-sen University}
  \city{ZhuHai}
  \country{China}
}

\author{Yuxin Su}
\email{suyx35@mail.sysu.edu.cn}
\authornote{corresponding author}
\affiliation{%
  \institution{Sun Yat-sen University}
  \city{ZhuHai}
  \country{China}
}

\author{Dan Li}
\email{lidan263@mail.sysu.edu.cn}
\affiliation{%
  \institution{Sun Yat-sen University}
  \city{ZhuHai}
  \country{China}
}

\author{Zibin Zheng}
\email{zhzibin@mail.sysu.edu.cn}
\affiliation{%
  \institution{Sun Yat-sen University}
  \city{ZhuHai}
  \country{China}
}

\renewcommand{\shortauthors}{Trovato et al.}

\begin{abstract}
Logs are essential for software systems to perform anomaly detection, but the scarcity of high-quality public log datasets has significantly hindered the development of log-based anomaly detection techniques. Existing public log datasets suffer from three fundamental limitations: (1) incomplete event coverage, (2) lack of authenticity in static analysis-based automated generation frameworks, and (3) insufficient semantic awareness. To address these challenges, this paper proposes AnomalyGen, the first automated log synthesis framework for anomaly detection. The framework features a novel four-phase architecture that synergistically combines\textbf{ enhanced program analysis} with \textbf{Chain-of-Thought} (CoT) reasoning to enable iterative log generation and anomaly annotation without requiring actual system execution. Experimental evaluations on Hadoop and HDFS distributed systems demonstrate that AnomalyGen achieves significantly higher log event coverage (\textbf{38×–95×} improvement over existing datasets) and generates more realistic log sequences compared to static analysis-based tools. Wh\-en augmenting benchmark datasets with AnomalyGen-generated data, we observe up to \textbf{3.7\%} improvement in F1-score (\textbf{1.8\%} average improvement across three state-of-the-art anomaly detection models). This work establishes a high-quality benchmarking resource for automated log analysis and pioneers a new paradigm for LLM applications in software engineering.

\end{abstract}





\setcopyright{none} 
\settopmatter{printacmref=false} 
\maketitle

\section{Introduction}

With the exponential growth in complexity of modern software systems \cite{zhu2023loghub}, runtime logs have become a core data source for software debugging and maintenance \cite{mastropaolo2022using}. Research has shown that fine-grained logging can not only pinpoint the root cause of program execution failure, \cite{mizouchi2019padla} but also significantly improve the accuracy of anomaly detection. 
However, the dual constraints of storage cost and performance overhead \cite{li2024logshrink} make long-term recording of complete log streams a major challenge in engineering practice \cite{mizouchi2019padla}. 
The current mainstream selective logging strategy relieves storage pressure but may lead to permanent loss of critical diagnostic information. This paradox has driven an explosion of research in the field of log mining, covering anomaly detection
\cite{li2020swisslog,yang2021plelog,du2017deeplog,huo2023semparser,liu2023scalable} and root cause analysis \cite{wang2021groot,lu2017log,wang2020root,chen2021pathidea,li2022intelligent}. 
Previous researchers have proposed innovative solutions such as semantic embedding-based similarity detection \cite{shavit2024semantilog}, improved PCA methods fusing control flow features \cite{yang2024try}, and so on. However, the practical efficacy of these methods is always limited by the severe lack of high-quality log datasets \cite{huo2023autolog,zhu2023loghub,jiang2024large}, which has become a central bottleneck constraining the development of the field.


The current mainstream methods for constructing log datasets can be divided into two categories:
1) Passive collection mode: Logs are directly obtained by monitoring running systems, such as LogHub \cite{zhu2023loghub} and LogPAI \cite{logpai}. Although this type of method can ensure the authenticity of the logs, it has the inherent defect of limited coverage and is very time-consuming (Collecting logs for Zookeeper takes more than 26 days, but only  207,820 messages)~\cite{he2020loghub,oliner2007supercomputers}.
2) Static generation mode: Automatic log generation methods based on program analysis, such as AutoLog \cite{huo2023autolog}. This method uses static guiding method to discover execution paths and generates reasonable log sequences \cite{yuan2012improving,zhao2017log20}. While such approache extends coverage, static analysis may lose many dynamic control flow features. For example, at real runtime different branches are decided based on input data, but static generation tools may fix the number of loops. Another example is that system reflection calls invoke different methods at runtime depending on the context, but static tools often only capture part of the call branch, leading to semantic distortion.



We identify three core challenges in the construction of current log datasets:

\textbf{The comprehensiveness of log events dilemma.}
Runtime logs struggle to iterate complex anomaly scenarios across the full system lifecycle~\cite{amar2019mining}. Current mainstream passive log collection methods rely on limited workloads: Hadoop log dataset \cite{lin2016log}, for example, contains only WordCount and PageRank test programs, and is designed to cover only the three basic types of anomalies, namely machine crashes, network outages, and disk full. However, the real-world anomaly scenarios are more varied than these simple simulations. There is an order of magnitude gap between the diversity of anomalies in the scenario \cite{wang2021tsagen}. 
    
\textbf{The veracity chasm of static analysis.} Automatic log generation tools based on static analysis (e.g., AutoLog\cite{huo2023autolog}) have inherent flaws: lack of runtime information, difficulty in mapping reflective calls, inability to automatically infer dynamic parameters such as the number of loops, lack of control when calling methods, and data-flow context checking and so on. All these things will lead to a lack of authenticity in the generated log sequences. In addition, logs generated by such tools also generally lack information on key parameters, significantly affecting the feature extraction capability of anomaly detection models. Busse et al.\cite{ref} Through fault injection experiments, it is confirmed that the trace information generated by static analysis has limited guidance in vulnerability localization, revealing the essential limitations of the static analysis approach.

\textbf{Semantic awareness capability is missing.} The log sequences generated by existing tools generally lack execution context semantics: the mapping between log records and code control flow is ambiguous, and DevOps needs to spend a lot of time tracing the mapping between logs and code control flow \cite{Gao1,Chen5}, and the Existing generated logs lack semantic support for execution path context, which significantly reduces the efficiency of exception localization and makes it difficult to trace the cause of exception triggering.




To address the above challenges, this paper proposes AnomalyGen - the first automatic log generation framework that incorporates large language modeling. The framework is to utilize enhanced program analysis as well as the LLM chain of thought to collaboratively iterate through log generation without dynamically executing logs. Specifically, AnomalyGen consists of the following four core phases:

The log-related call graph construction phase first extracts the global call graph through static program analysis, and prunes the log-irrelevant nodes using domain knowledge. The fine-grained enhanced static analysis phase obtains the local subgraph by double threshold setting, and uses LLM to convert the node's ternary information \textbf{< Source Code, Call Path, Log-Oriented CFG >} to generate the single node's Enhanced\_CFG. The CoT-based recursive log sequence generation phase utilizes the execution graph to recursively integrate the sub-node sequences in accordance with the bottom-up strategy, and utilizes the CoT capability of the LLM to logically verify the data flow and control flow during the merging process. Anomaly annotation phase based on knowledge rules. Combined with domain expert knowledge, the log sequences are filtered and annotated by explicit and implicit rules, and ultimately high-quality data containing real anomaly scenarios are generated. 

We completed the above experiments on two popular large-scale distributed systems, Hadoop and HDFS, and demonstrated the superiority of AnomalyGen from four perspectives. (1) The experiments show that AnomalyGen acquires many more log events on the same system than existing datasets (5,094 more logs than the benchmark Hadoop dataset, and 95.3 times more than the benchmark HDFS dataset), and that the fidelity of the log sequences outperforms that of other tools that perform log generation based on static analysis (the average comprehensive coverage of log events is 97.48 \%) (2) By comparing the existing advanced log generation tools through example analysis, we find that AnomalyGen adds the inference verification of LLM CoT and parameter simulation to the automatic log sequence generation framework based on program analysis only, and it can deal with part of the logs of dynamic calls and identify more anomalous scenarios than the other tools to improve the authenticity of the log sequences. (3) In the process of simulating and generating logs, we utilize the semantic sensing ability of the big model to retain the corresponding execution paths and corresponding condition values of the logs, which mitigates the defects in control flow caused by the lack of semantics in the static analysis context, and makes the log sequences more traceable for the maintenance personnel to locate the anomalous logs and trace the log execution flow quickly. (4)  When augmenting benchmark datasets with AnomalyGen-generated data, we observe up to 3.7 \% improvement in F1-score (1.8 \% average improvement across three state-of-the-art anomaly detection models). AnomalyGen has proven to provide higher quality datasets for anomaly detection techniques.



In summary, the contribution of this paper is as follows:

\begin{itemize}
    \item We present AnomalyGen, a novel and widely applicable approach to automated log generation using LLM, which addresses three limitations of existing log datasets: lack of comprehensiveness, veracity, and semantic-awareness capabilities.
    \item AnomalyGen has four phases: finding log-related pruning graphs, generating single-node augmented CFGs, recursive log merging based on CoT validation, and anomaly labeling based on knowledge rules.
    \item Experimental results on two popular large-scale distributed systems, Hadoop and HDFS, show that AnomalyGen achie\-ves comprehensive coverage of log events (\textbf{97.48} \%) and generates a more realistic log dataset, and we further demonstrate that when augmenting benchmark datasets with Ano\-malyGen-generated data, we observe up to \textbf{3.7 \% improvement} in F1-score (\textbf{1.8 \% average improvement} across three state-of-the-art anomaly detection models). 
    \item To the best of our knowledge, AnomalyGen is \textbf{the first approach to automated log generation using LLM for anomaly detection}. All artifacts and datasets have been released for future research.
\end{itemize}


\begin{table*}[!htb]
    \caption{Statistics and descriptions of existing datasets.}
    \label{tab:1}
    \centering
        \begin{tabular}{c|ccccccc|ccccccc}
        \hline
            & \# Log Event & \# Workload & \# Failure Type & \# Message & Collection Time & Granularity\\ 
        \hline
        R-Hadoop & 242 & 2 & 3   & 394,208 & NA & Application andContainer \\ 
        R-HDFS & 30 & NA & 11 & 11,175,629 & 38.7 Hours & Block\\ 
        \hline
        \end{tabular}
\end{table*}

\section{Motivation Study}
\subsection{Study Subject}
By systematically analyzing the log-based anomaly detection papers published in the last three years (68 \% of CCF-A conferences), we select two widely used distributed system log datasets as the study subjects (Table \ref{tab:1}). D-Hadoop \cite{lin2016log}, D-HDFS \cite{xu2009detecting} dataset collected by injecting predefined faults during normal workloads, with anomaly annotation. \cite{xu2009detecting}datasets are collected by injecting predefined faults in normal workloads with anomaly annotation (Hadoop with Application granularity, HDFS with Block granularity). As benchmark datasets in the current domain, their limitations can typify the common problems of existing log generation techniques.
Through case studies represented by them, we validate the following three key challenges with current datasets.


\subsection{Are existing datasets comprehensive?}
There is a significant event coverage gap in the existing dataset. The Hadoop benchmark dataset \cite{lin2016log}, for example, contains only two test programs, WordCount and PageRank, and the types of injected anomalies are limited to machine crashes, network misconfigurations, and disk overloads. This narrow coverage makes it difficult to capture complex combinations of anomalies in real systems (e.g., the superimposed effects of concurrent resource competition and network jitter). This passive collection approach also leads to a contradiction between data quality and scale -- to increase the anomaly rate from 0.1 \% to 15 \%, 150 times the normal log data would need to be deleted \cite{le2022log}and the combinatorial explosion characteristic of the dynamic execution path is even more important. The combinatorial explosion nature of dynamic execution paths makes it even more difficult to trigger deep anomalies with traditional testing methods, and the cost of testing is growing super-linearly. More critically, the anomaly patterns found by the AutoLog\cite{huo2023autolog} tool in the D-HDFS dataset suggest that existing methods may ignore log sequences related to responders' connection timeouts, resulting in a limited ability to generalize the anomaly detection model to unknown scenarios. \cite{cinque2010assessing,pecchia2012detection}As a result, anomaly detection techniques trained using incomplete datasets may not be able to handle unseen logs \cite{cinque2010assessing,pecchia2012detection}, and would make the experimental conclusions potentially unrepresentative. This is because these datasets lack comprehensive coverage and have significant gaps with real-world scenarios\cite{huo2023autolog}.


\begin{figure}[h]
\centering
\includegraphics[width=0.9\linewidth]{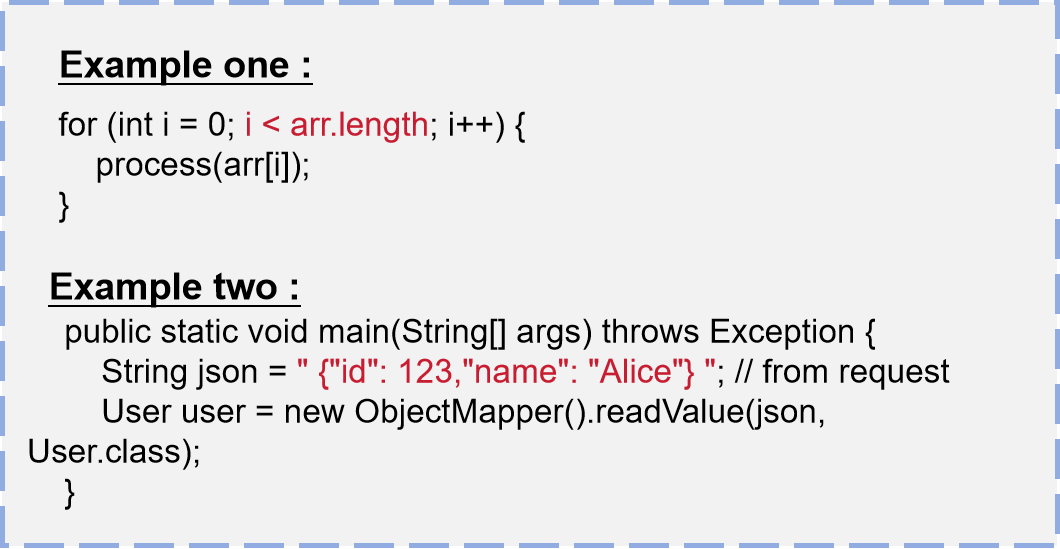}
\caption{Two Code Cases That Are Difficult to Handle with Static Analysis}
\label{static}
\end{figure}

\subsection{Can existing analysis tools generate real logs?}

 Although static analysis tools enable automatic log generation, they have significant limitations in capturing dynamic runtime characteristics. For example, in reflection scenarios such as a common function Class.forName(), static analysis is unable to resolve dynamic method calls and variable bindings, resulting in missing parameters in the generated logs. AutoLog \cite {huo2023autolog} attempts to simulate loop iterations through artificial constraints, but the lack of runtime context produces unrealistic control flow Transfer. Take a simple loop as an example, as shown in Figure \ref{static} Example one.Static tools may expand loops with arbitrary values (e.g. i=0,1,2) instead of using the actual array length at runtime, resulting in incomplete or incorrect logs.
 
More seriously, static analysis is fundamentally flawed when dealing with dynamic variable dependencies. When a program constructs complex objects by receiving data over the network or user input.For exmample, as shown in Figure \ref{static} Example two. Static analysis is unable to infer the specific structure and value range of data, resulting in null values or default placeholders in the logs. Experiments by Busse et al.\cite{busse2022combining} show that the search effectiveness of such unrealistic logs in guiding vulnerability mining is only 37 \% of that of real logs. A typical case is buffer overflow vulnerability detection, where logs generated by static tools often ignore dynamic changes in heap memory allocation, resulting in an inability to reproduce boundary conditions similar to those in the Heartbleed vulnerability.
These limitations stem from the inherent flaw that static analysis cannot model runtime state and environment interactions. While dynamic analysis can capture real logs, it suffers from high overhead and scalability challenges. Therefore, there is a need to combine the structural comprehension capabilities of static analysis with the context-aware parameterization of dynamic sampling to build hybrid approaches to bridge the semantic gap in log generation. 



\subsection{Do existing tools support contextual semantics?} 
Current log generation tools lack the capability to preserve execution context, posing significant challenges for DevOps troubleshooting. In many cases, most of the root cause analysis effort is spent manually correlating logs with the corresponding code control flow. For example, our replication using the D-Hadoop dataset revealed that the HDFS NameNode failure logs were missing critical call path details—such as the replication pipeline state—which are typically maintained in files like \textit{hdfs.server.namenode.ReplicationMana\-ger.java} and \textit{hdfs.server.namenode.FSNamesystem.java}. As a result, experimenters had to inspect more than 15 separate source files for each exception, severely impeding troubleshooting efficiency.

The inability of existing tools to preserve dataflow and controlfl\-ow dependencies directly restricts the application of automated root cause analysis techniques, while also limiting the anomaly detection domain from effectively leveraging contextual information.

\subsection{summary}
The above findings reveal three major requirements for next-gener\-ation log generation technologies:
\begin{itemize}
    \item Comprehensive event coverage through proactive exploration of system state.
    \item Preserve contextual semantics, inferred control flow and data flow through dynamic synthesis
    \item Generate traceable log sequences to support accurate anomaly localization through context retention.
\end{itemize}

Our solution--AnomalyGen, meets these requirements through program analysis enhanced by a Large Language Model (LLM) and iterative verification based on the Chain of Thought. The details are presented in section \ref{Methodology}, section \ref{Implementation} and section \ref{Experiments}.  


\section{Methodology}\label{Methodology}

\begin{figure*}
  \includegraphics[width=\textwidth]{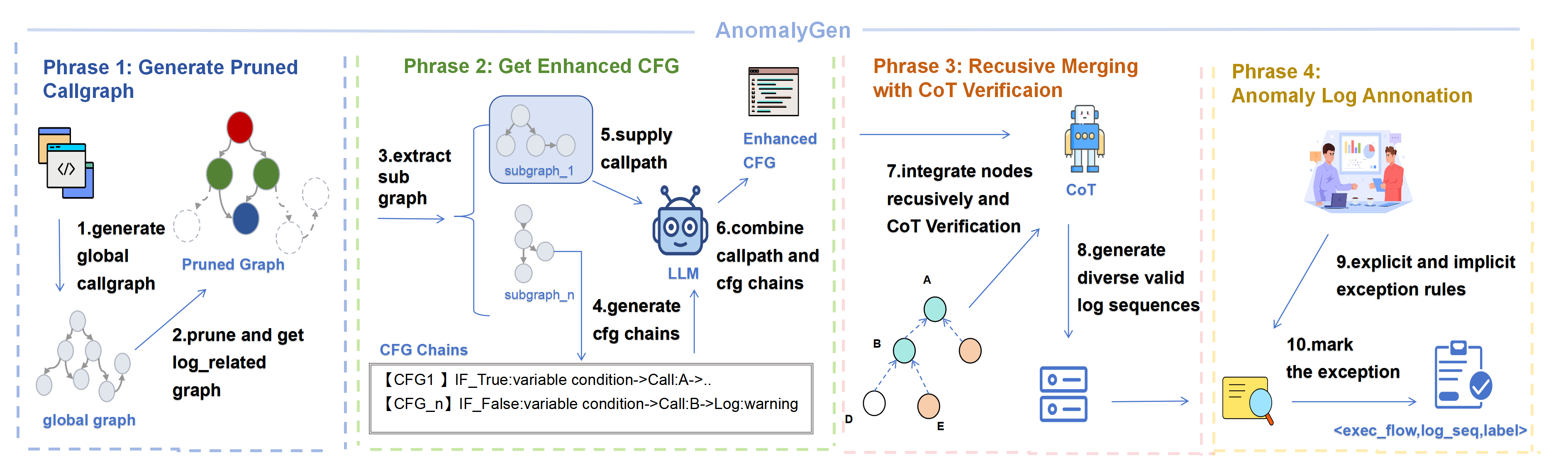}
  \caption{The Four Satges of AnomalyGen Framework.}
  \label{fig:Framework}
\end{figure*}


\subsection{Overview}
Existing log anomaly detection techniques are often limited by the comprehensiveness and authenticity of log data, and automatically generating log sequences as a training dataset provides a feasible solution to the problem of the lack of high-quality datasets. The core challenge of automatically generating log sequences is how to accurately extract execution paths from a large code base and generate semantically coherent log sequences. Existing approaches are often caught in a dilemma: over-reliance on manual rules leads to limited coverage, or simple application of static analysis leads to problems such as path explosion, unrealistic sequences generated by simulation.

Existing log anomaly detection techniques are constrained by the comprehensiveness and authenticity of log data. Automatically generating log sequences as training datasets provides a viable solution to the scarcity of high-quality datasets. The core challenge lies in accurately extracting execution paths from large codebases and generating semantically coherent log sequences. Existing approaches often face a dilemma: over-reliance on manual rules leads to limited coverage, while simplistic static analysis results in path explosion or unrealistic sequences.
To address these issues, AnomalyGen employs a four-stage progressive enhancement framework. Firstly,construct accurate log call subgraphs via static analysis. Secondly, conduct fine-grained program semantic analysis. Thirdly, merge recursively with LLM reasoning verification. Finally, apply knowledge rules for anomaly labeling. Applying our data to benchmark datasets, we observe up to 3.7 \% improvement in F1-score (1.8 \% average improvement across three state-of-the-art anomaly detection models).  Experimental results demonstrate significant performance improvements for log anomaly detection models trained on our generated datasets.

\subsection{PHASE I: Logging-related call graph pruning}

\begin{figure}[!htb]
    \centering
    \includegraphics[width=0.4\textwidth]{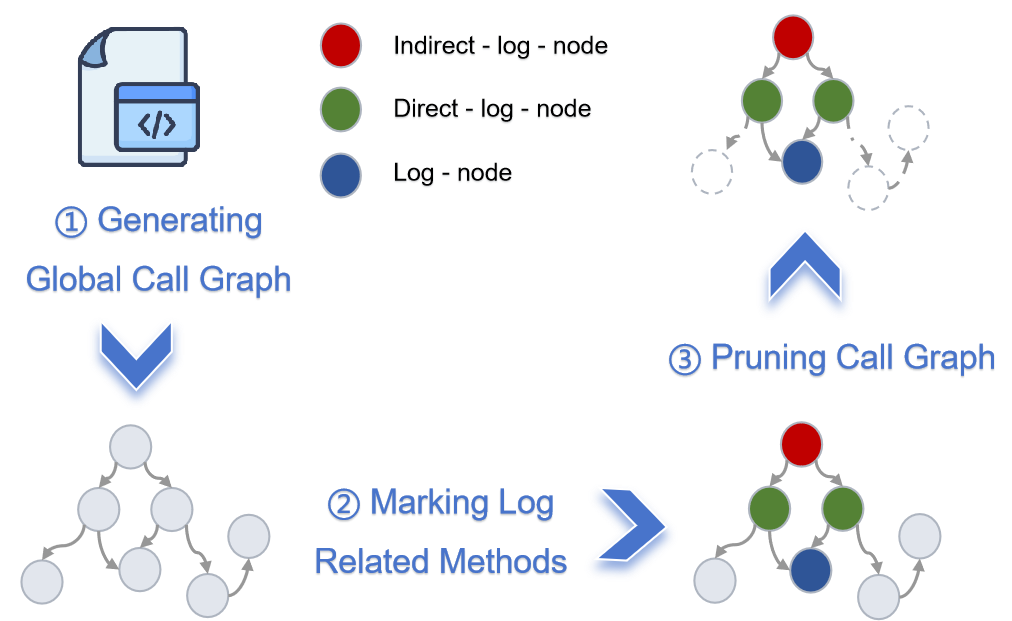}
    \caption{ PHASE I: Log-Related Node Labeling and Pruning.}
    \label{phrase_1} 
\end{figure}

Analysis of Hadoop 3.3.6 source code reveals its call graph contains 295,216 method nodes, with only 1.91\% (5,635) directly invoking logging APIs. Blindly traversing the entire graph leads to computational waste and noise interference. We address this with a two-step pruning approach:

    \subsubsection{Global Call Graph Generation}

    In order to get the log-related call nodes, we use the static analysis tool to generate a complete cross-method call graph based on the entire repository. Static analysis tool generates a complete cross-method call graph, where nodes represent functions and edges denote call relationships (caller→ca\-llee). For Hadoop, this results in 295,216 node\-s and over 980,000 edges. With the aim of focusing the analysis on log-related content, we need to prune on this large graph.


    \subsubsection{Log Node Tagging and Reverse Pruning}
    
\begin{figure*}[!htb]
    \centering
    \includegraphics[width=1\textwidth]{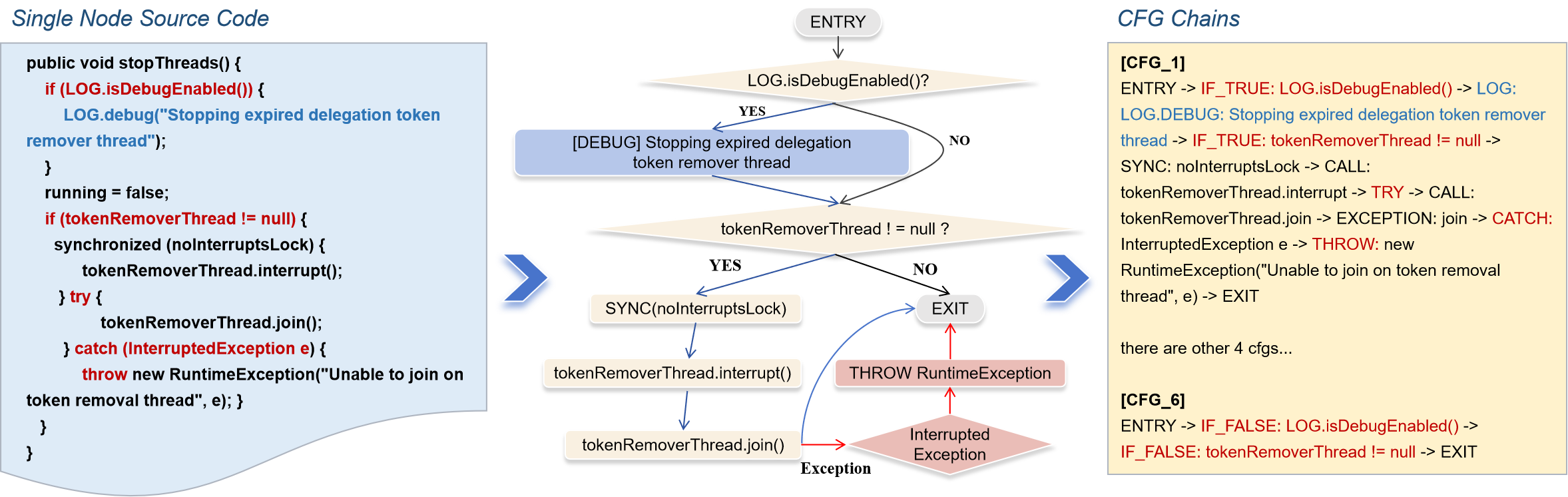}
    \caption{Log-Oriented Control Flow Graph and Control Chain Generation.}
    \label{fig:RQ2}
\end{figure*}

In large call graphs, performing a forward traversal to extract all edges and then applying log-related pruning can lead to excessive computational overhead and path explosion \cite{nejmeh1988npath}. To overcome this, we use logging API calls as anchor points and trace the call chain in reverse, progressively marking methods related to logging.
Specifically, a node can generate logs in two ways: either by directly generating logs or indirectly through its child or grandchild nodes. Accordingly, we define a node that directly calls a logging framework API(e.g., org.slf4j.logger:info \cite{slf4j2022simple}) as a Direct\_Log\_Method; any node that calls or indirectly calls a Direct\_Log\_Method is defined as an Indirect\_Log\_Method. To efficiently label these nodes, we first summarize common logging frameworks (with AnomalyGen supporting extensibility via configuration), then, after identifying the Direct\_Log\_Method nodes, we recursively traverse upward to mark all Indirect\_Log\_Method nodes. Finally, we prune unmarked nodes and their corresponding edges. As shown in Figure \ref{phrase_1}, the process begins with generating a global call graph from the code, followed by marking and pruning. In the figure, blue nodes represent log nodes, green nodes indicate Direct\_Log\_Method (i.e., nodes that directly output logs), while Indirect\_Log\_Method nodes are those that directly or indirectly invoke a logging output node.

This pruning process results in a significantly smaller pruned graph compared to the original call graph, as most methods do not include log statements or involve any log-related calls. For example, after pruning the call graph in the Hadoop system, only 15.11 \% (44,593 out of 295,216) of the nodes remain. Since only the nodes in the pruned graph are used for further analysis, this step greatly enhances the efficiency of AnomalyGen and facilitates subsequent analysis.
    

\subsection{PHASE II: Fine-grained Subgraph Mining and Control Flow Diagram Enhancement}

\begin{figure}[!htb]
    \centering
    \includegraphics[width=0.45\textwidth]{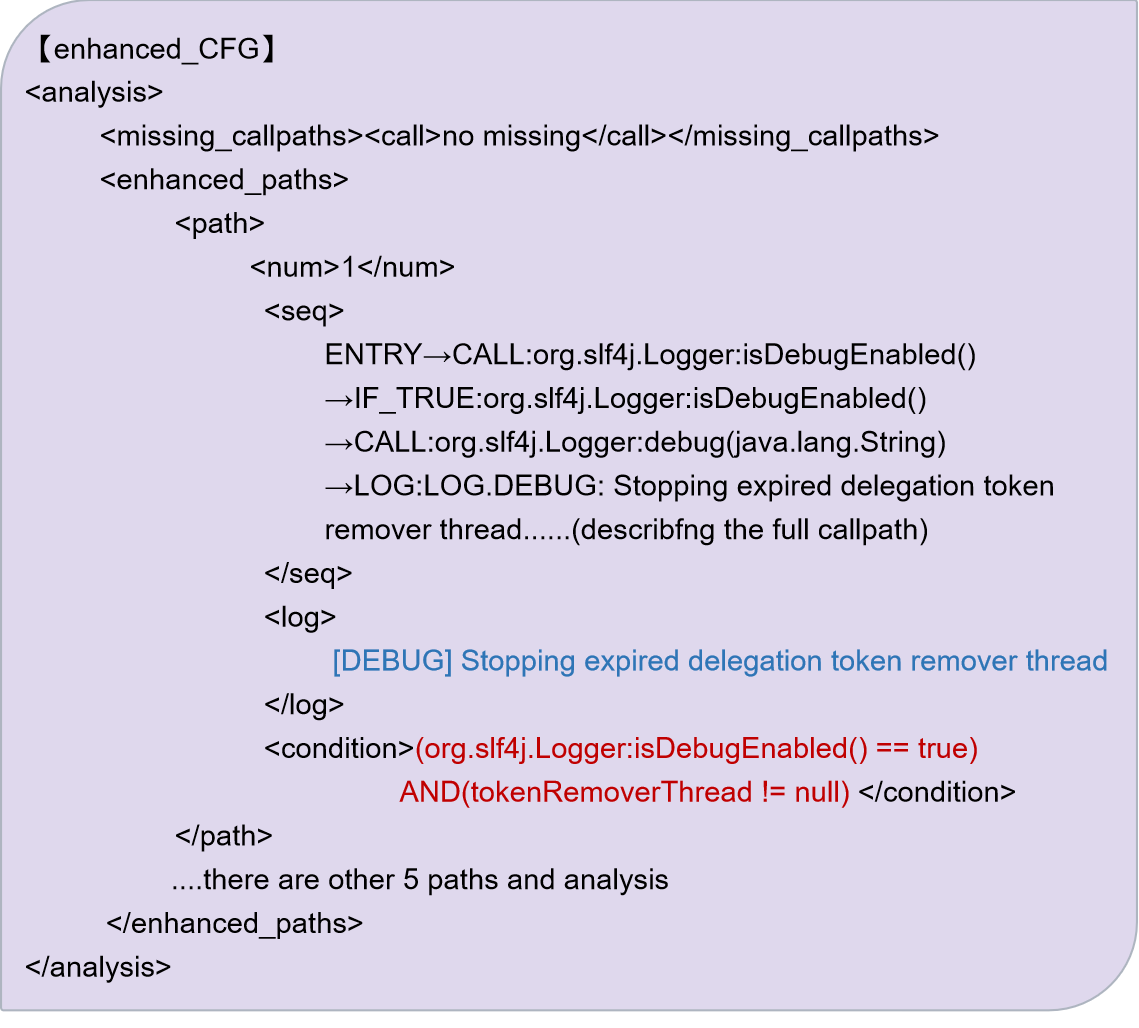}
    \caption{Structured Enhanced\_CFG Example.}
    \label{fig:RQ2-1}
\end{figure}

Although the pruned global call graph is focused on logging, directly enumerating all execution paths still runs the risk of path explosion. Due to the above limitation, we need to limit the scope of analysis to prevent redundancy of information. At the same time, we also need to obtain fine-grained contextual information in order to obtain fine-grained control information about the execution process and thus connect the execution flow. To balance these two requirements, AnomalyGen extracts moderate-sized subgraphs and uses static analysis techniques to capture the fine-grained relationships between control flow and calls preparing for recursive joining of nodes.


\subsubsection{Dual-threshold Subgraph Extraction}
Set an entry threshold and a depth threshold to extract representative subgraphs from the global graph that each contain at least one log statement. This strategy preserves key log points while avoiding path explosion. By partitioning the large graph, each segment can be analyzed more precisely, which complements later LLM inference—especially given that our experiments show LLMs struggle with long log sequence dependencies. Therefore, subgraph slicing is essential.


\subsubsection{ Log-Oriented Control Flow Graph (LCFG)}
To address two critical challenges—mapping branch conditions to log sequences and capturing log context in nested calls—we employ a layered analysis approach, focusing on solving the following two core problems: 
(1) Same-layer control flow analysis: for the branching structure (e.g., if/else, loops, switches, etc.) in the execution of a program, the mapping relationship between the branching conditions and the log output sequences is established through path-sensitive analysis;
(2) Cross-Level Call Analysis: Based on the dynamic characteristics of function call stack, capture the log context correlation during deep-level calls, which is manifested in the following ways: tracking function call/return sequences through call instructions, parsing nested hierarchical relationships of log call points, and establishing timing constraints on cross-level log events;
Based on the above analysis we realize the following three steps at the technical level:
a. Basic element extraction: Based on method granularity static code analysis techniques, identify the following key elements: control flow structure (branch/loop statements); function call points (call instructions); log output points (including log level call statements)
b. Rule-driven annotation system: Design formal annotation rulesets, including: control flow path constraint annotation, log call context annotation, cross-method call chain annotation.
c. Graph Structure Generation: Construct method-level LCFGs through combinatorial analysis, whose features include: displaying all feasible execution paths, visualizing the log output sequence of each path, and labeling the causal relationships between key control nodes and logs.

The graph structure provides fine-grained path constraints and log context dependencies for subsequent anomaly pattern detection, laying the foundation for subsequent refined analysis.
  
  
  

\begin{figure*}[!htb]
    \centering
    \includegraphics[width=0.9\textwidth]{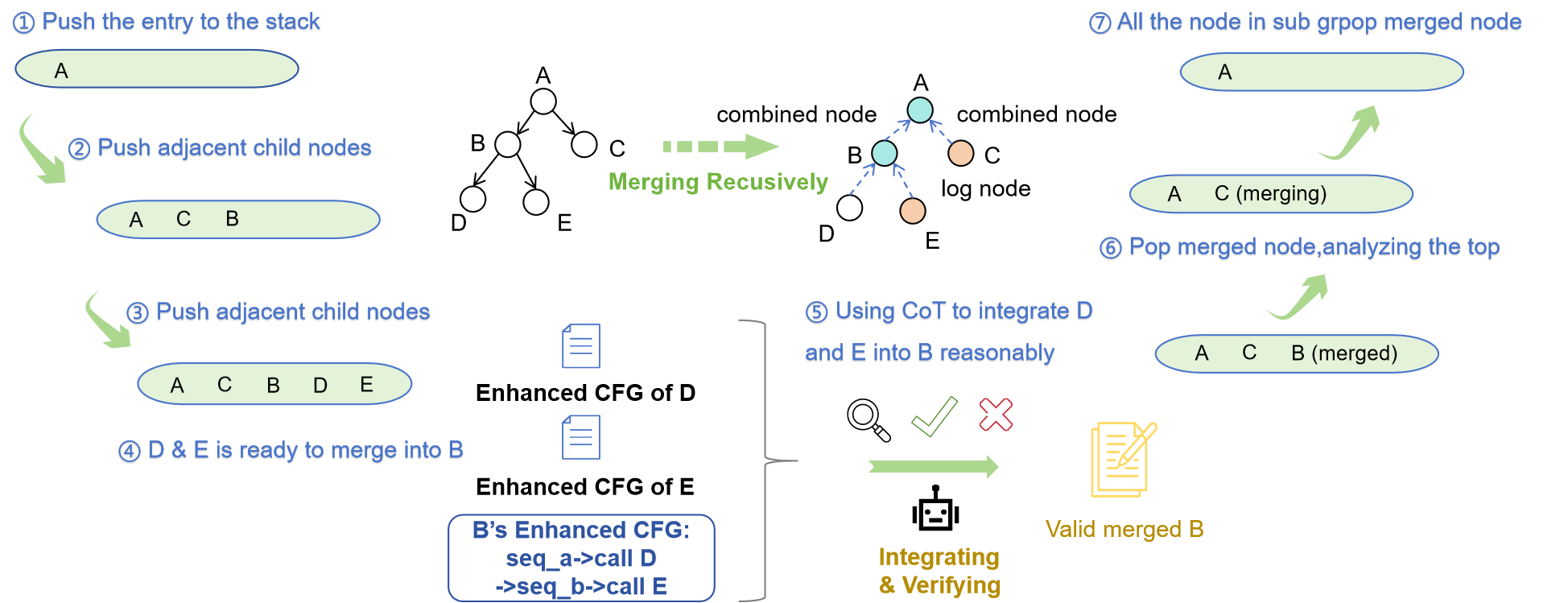}
    \caption{ PHASE III: Stack-based Bottom-up Recursive Log Merging and CoT Verification Process.}
    \label{fig:RQ2-2}
\end{figure*}

  \subsubsection{Enhanced CFG Generation Based on Ternary Information}
  Traditional static analysis struggles to accurately model Java's dynamic execution paths, especially when dealing with polymorphism and reflection. To address the limitations of the initial control flow graph (CFG) in handling complex calls, we propose an enhanced CFG generation method that incorporates call path information from bytecode analysis and LLM’s ability to infer implicit calls.Our approach constructs a ternary tuple: \textbf{< Source Code, Call Path, Log-Oriented CFG >}. By leveraging context awareness and inference capabilities of the LLM, we refine the CFG through a structured five-step verification process:
(1) Call Matching and Completion
Using cross-validation, we match method calls in the call path with CALL nodes in the CFG. For unmatched calls, we analyze the source code context and insert them into appropriate CFG positions (e.g., return or exception nodes).
(2) Exception Path Augmentation
By analyzing basic control structures (e.g., if, for) and cross-method call nodes, we identify and supplement potential exception-throwing nodes, refining both control flow and exception propagation paths.
(3) Log-Flow Association via LLM Reasoning
We utilize LLM reasoning to establish associations between log statements and control flow nodes, enabling variable tracking across dynamic execution scopes.
(4) Path Constraints and Exception Handling
For conditional expressions containing method calls, we retain invocation states to generate path constraints and introduce exception-triggering branches for nodes likely to throw exceptions.
(5) Execution-Log Consistency Verification
Through path reachability validation and log integrity checking, we ensure a one-to-one correspondence between execution flow and log sequences, guaranteeing that the enhanced CFG aligns with actual runtime logic.

This enhanced CFG provides a more precise representation of execution logic, offering richer contextual insights that improve log sequence reconstruction and anomaly detection.



\subsection{PHASE III:  Recursive Log Merging with CoT Inference Verification}

    \subsubsection{Data Structure of Analysis Process}
In practice, function calls follow a depth-first strategy, and log outputs frequently span multiple call levels—even involving nested loops. Analyzing logs function by function often leads to fragmented execution paths and loss of continuity. To address this, we simulate the "in-stack-out-stack" process using a stack structure to recursively merge log information from the bottom up. However, a simple merge can overlook constraints from conditions, control flows, and data flow conflicts. To mitigate this, we employ the CoT reasoning of a LLM to validate the merged sequences, reducing the risk of generating log paths that do not reflect actual execution.
  

\subsubsection{ CoT Path Reasoning Verification}
To ensure that the merged log sequences are both accurate and traceable, we instructed the larger model to evaluate and prune out irrelevant or invalid paths from multiple perspectives such as control-flow consistency, data-flow integrity, and call-point correlation. CoT is an important methodology that is frequently used to guide LLMs through complex reasoning in prompt prompt engineering, and so we utilized the LLM CoT to complete the reasoning as described below, Verification and generation, including the following three processes:
(1) AnomalyGen first identifies key log statements in parent and child node paths to construct a log fingerprint: <class name:method name> [log level]. Then, it performs call-point correlation localization to establish the mapping between parent method call stacks and child method entries, and retains only the child paths that contain valid log information.
(2) Exact Condition Fusion: The extracted call paths and log sequences are then checked to exclude unreasonable sequences and add dynamic variable information. AnomalyGen extracts dynamic variables from log statements to form a delivery chain, performs conditional conflict and data flow consistency detection, flags invalid sequences, and finally imposes constraints.
(3) Log Sequence Optimization: Finally, the merged sequences are further optimized by pruning strategy and log integrity checking to generate high-quality log output.

\begin{figure}[h]
\centering
\includegraphics[width=0.7\linewidth]{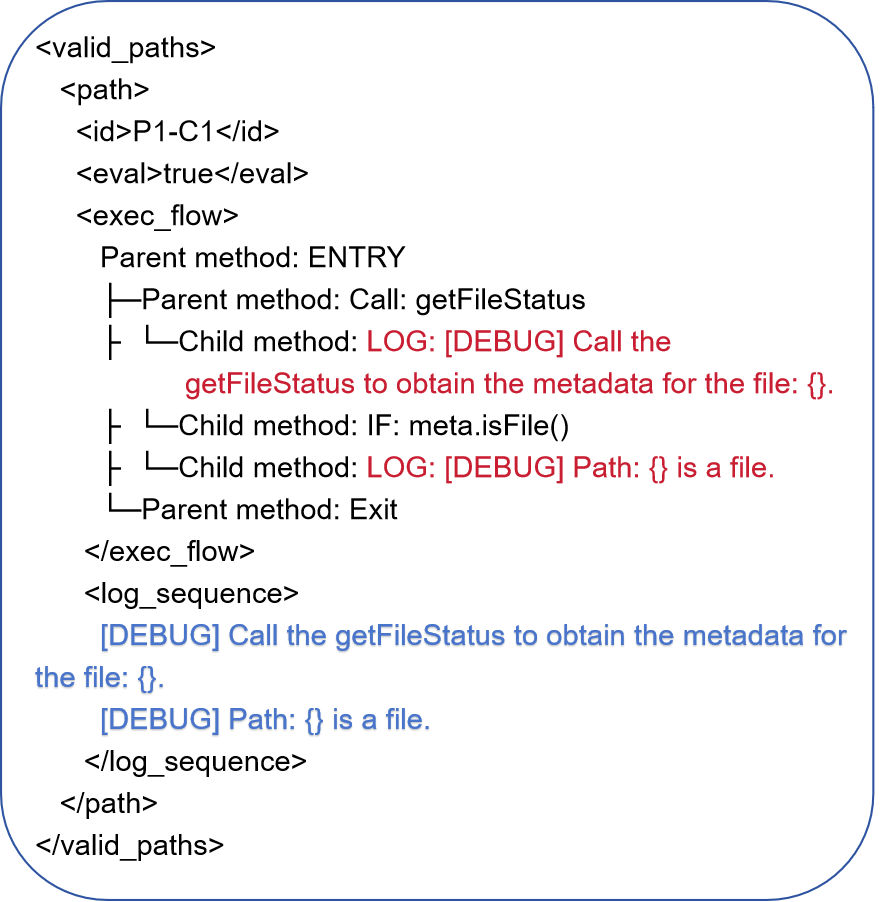}
\caption{Structured Output Validated by CoT.}
\end{figure}

This phase successfully integrates the log information in multi-layer function calls through recursive merging and chained reasoning verification, reconstructing the log sequences that match the actual execution flow, providing a complete semantic chain for anomaly detection.

        
\subsection{PHASE IV:Knowledge-Driven Exception Log Labeling}
There are two main challenges in annotating exception logs: on one hand, explicit exception hints (e.g., "Exception", "ERROR") are easy to capture; however, some exception information exists only as implicit signals, such as the presence of error codes or failure keywords. To face the above challenges, we explore how to capture explicit anomalies while identifying more potential anomalies through implicit features and introducing expert validation to ensure annotation quality.
This strategy has become a standard practice in mainstream logging frameworks (e.g., Log4j, SLF4J, etc.), which helps to quickly capture explicit exception information. Implicit exception reasoning focuses on identifying error codes (e.g., error\_code=500) as well as keywords describing failures (e.g., "fail", "cannot", "invalid"), these implicit signals capture problems that are not directly presented as exception keywords. Based on the above analysis, we specify the above annotation rules for both implicit and explicit methods for annotating the anomaly log sequences.
In order to verify the effectiveness of the above annotation rules, the experiments were conducted by three research generators who are also engaged in log-related research to randomly select 10 \% of the automatic annotation results for domain expert review, and to ensure the annotation consistency by Krippendorff's α \>\= 0.8, so as to further rectify possible omission or mislabeling problems. Finally we label each log sequence with its corresponding execution context and anomalies. The formation of <log\_sequence, execution\_context, anomaly\_label> constitutes the dataset, which provides high-quality training data for subsequent anomaly detection algorithms.


\section{Implementation}\label{Implementation}
AnomalyGen is primarily implemented using Java and Python. We leverage java-callgraph2 to generate global callgraph, javaparser to perform comprehensive program analysis and employ py4j to seamlessly transfer the parsing results to Python. To manage the extensive call path data, we utilize python-mysql for robust storage. Additionally, we integrate large language models such as GPT-4o and DeepSeek-Chat to facilitate chain-of-thought (CoT) reasoning. Our experiments are conducted on Ubuntu 22.04 with a 13th Gen Intel(R) Core(TM) and Windows11 with 12th Gen Intel(R) Core(TM).

\section{Experiments}\label{Experiments}
The purpose of this part of the experiment is to verify the comprehensiveness of AnomalyGen in log dataset generation, the authenticity of recursively merged logs, the contextual semantic awareness capability, and its practical gain for anomaly detection problems. The experimental design fully corresponds to the four major requirements of the Motivating Study, and demonstrates the advantages of AnomalyGen in real-world scenarios through quantitative metrics and comparative analysis,they are:
\begin{itemize}
    \item RQ1: How comprehensive is the data generated by AnomalyGen?
    \item RQ2: Can AnomalyGen generate more realistic logs than existing methods?
    \item RQ3: Can AnomalyGen retain contextual semantic information?
    \item RQ4: Can AnomalyGen benefit anomaly detection problems?
\end{itemize}


\subsection { Experimental Settings}
1) Baseline dataset: The widely used D-Hadoop and D-HDFS datasets are used, which are injected with multiple faults under normal workloads and contain detailed anomaly annotations.
D-Hadoop~\cite{lin2016log}. Hadoop is an open source framework designed to efficiently store and process large-scale data. The dataset is collected by running two standard applications and injecting three types of faults for anomaly detection.
D-HDFS~\cite{xu2009detecting}. HDFS is a distributed file system for big data storage that enables High-throughput access to data. D-HDFS is collected from private cloud environments executing benchmark workloads with labeled anomalies.
2) Metrics: We follow the various metrics in AutoLog for evaluating the comprehensiveness of the coverage of the log dataset. This includes the total number of log events, coverage($\frac{\#Log\_Event}{\#Total\_Log\_Event}$), and the effect of coverage on existing datasets ($\frac{\#Log\_Event\_coverage}{\#Log\_Event\_in\_Existing\_Dataset}$).

\begin{table*}[!htb]
    \caption{The comparison of datasets for comprehensiveness. R-coverage is reported for systems with publicly log datasets.}
    \label{tab:2}
    \centering
        \begin{tabular}{c|ccccccc|ccccccc}
        \hline
            &Dataset & \# Log Events & \# Logging Coverage & \# R-Coverage & Increment\\ 
        \hline
        Hadoop & R-Hadoop & 242 & 2.50\%(242/9662)   & - & 38X \\ 
        - & LHub-Hadoop & 107 & 1.11\%(107/9662)   & 86.92\%(93/107) & 86X \\ 
        - & AG-Hadoop & 9225 & 95.48\%(9225/9662)   & - & - \\
        HDFS & R-HDFS & 30 & 1.04\%(30/2889)   & - & 95X \\
        - & LHub-HDFS & 15 & 0.52\%(15/2889)   & 93.33\%(14/15) & 191X \\
        - & AG-HDFS & 2874 & 99.48\%(2874/2889)   & - & - \\
        \hline
        \end{tabular}
\end{table*}

\subsection {RQ1: How comprehensive is the data generated by AnomalyGen?}
We evaluated the comprehensiveness of the dataset generated by AnomalyGen in terms of the number of log events, coverage of log records, and coverage compared to the baseline dataset, as shown in the table. Compared to existing logging datasets, AnomalyGen effectively enhances the comprehensiveness of all the projects' logging events, achieving an average coverage of about 97.48 \%, ensuring that more than 97 \% of the designed logging events are covered. The number of logging events generated by AnomalyGen is 38 to 95 times higher than that of traditional datasets (e.g., HDFS). The number of log events generated by AnomalyGen is 95 times higher than that of traditional datasets (e.g., the HDFS scenario scales 95 times from 30 to 2874 events, and the number of logs in the Hadoop system scales 37 times from 242 to 9225 log times). Regarding the missing parts, we analyze them as follows: their limitation mainly stems from the limited construction of call graphs, since the analysis of AnomalyGen's initial call graphs is based on static analysis, and even though the absence of dynamic variables and executions is partially bridged by the semantic-awareness capabilities of the larger model, there are still complex dynamic reflections that are difficult to analyze, as well as missed analyses due to variable uncertainty. In addition, we conduct a detailed validation analysis on a smaller volume of dataset log events with manual validation, and the experimental results show that AnomalyGen covers most of the datasets in the existing data, with a ratio of 93/107 for Hadoop\_2k(LHub-Hadoop) and 14/15(LHub-HDFS) for HDFS, respectively. (This is slightly different from the total data logged by loghub, as the above experiments are all completed based on hadoop v3.3.6, we partially filtered the dataset because of log modifications due to version changes) achieved an average R-coverage of 90.125\%, which is better than the existing log generation framework.


AnomalyGen achieves an average coverage rate of 97.48\%, which is 37 to 94 times higher than that of log events in traditional datasets, fully demonstrating that AnomalyGen effectively enhances the comprehensiveness of log events and provides a richer sample base for subsequent anomaly detection.

\subsection{RQ2: Can AnomalyGen generate more realistic logs than existing methods?}  
To evaluate the authenticity of logs generated by AnomalyGen, we compare it with existing tools (e.g., AutoLog) in three dimensions: dynamic logging method parsing, control flow integrity, and dynamic parameter simulation, and validate its advantages through code examples.

\begin{figure}[h] 
\centering
\includegraphics[width=0.8\linewidth]{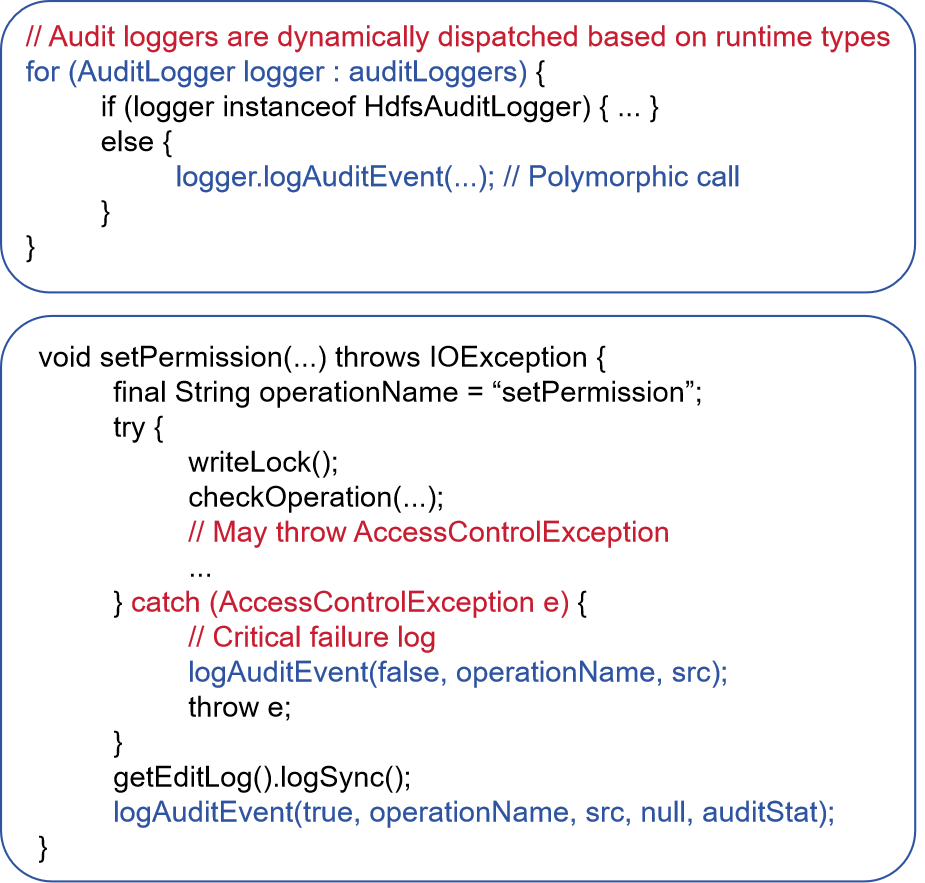}
\caption{Example Code with Polymorphic Calls and Exception Branches.}
\label{ex1}
\end{figure}

\begin{figure}[h]
\centering
\includegraphics[width=0.8\linewidth]{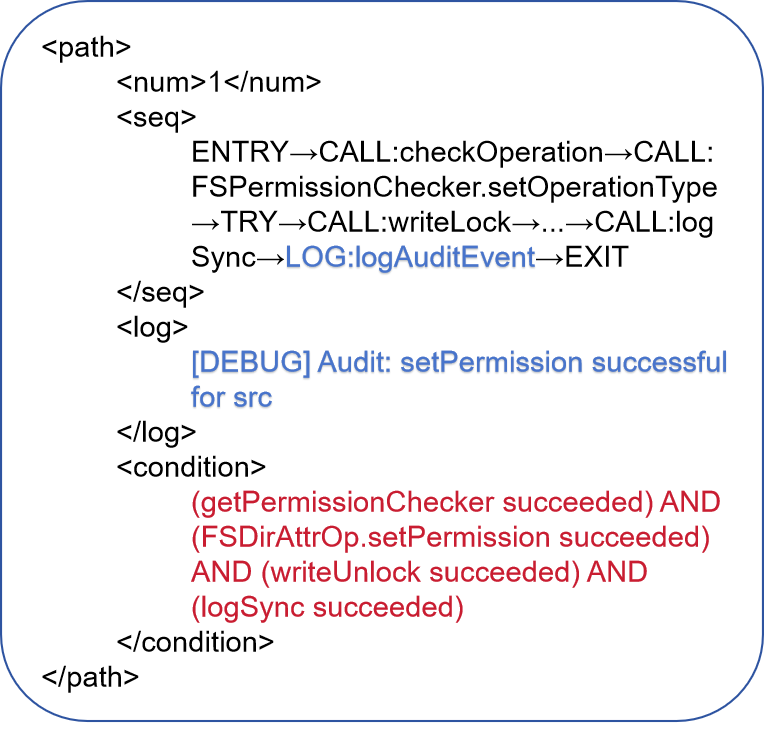}
\caption{CFG Diagram of the Exception Normal Flow for Function (setPermission)}
\label{normal_flow}
\end{figure}

\begin{figure}[h] 
\centering
\includegraphics[width=0.8\linewidth]{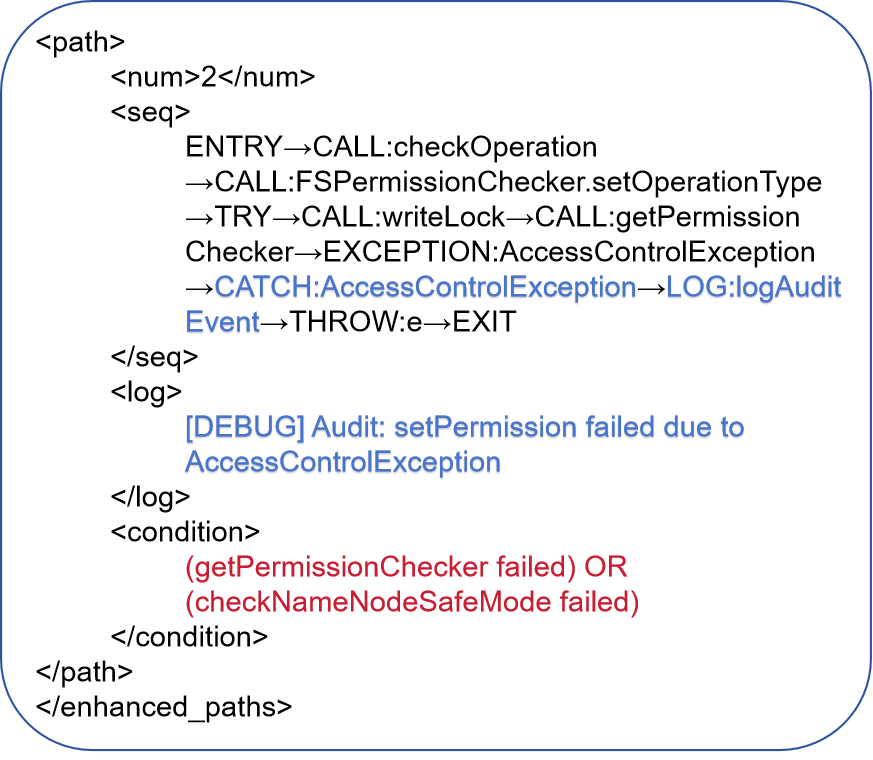}
\caption{CFG Diagram of the Exception Control Flow for Function (setPermission)}
\label{exc_flow}
\end{figure}

(1) Omission of Dynamic Invocation Log Method 
Existing tools (e.g., AutoLog) rely on static analysis to identify logging invocations, and are unable to resolve logging methods triggered by reflection or polymorphic mechanisms. For example, in the HDFS audit logging code \ref{ex1}: the actual type of AuditLogger in the above figure cannot be obtained in the static analysis, which belongs to the dynamic polymorphic invocation, and AutoLog can only detect the static declaration of logAuditEvent in the figure below, but cannot infer the specific implementation class, and thus cannot get the content of the audit log. But AnomalyGen through the call path tracing with LLM enhanced contextual reasoning: through iterative verification to rebuild the polymorphic call paths, through the semantic perception of context, simulation to infer the log corresponding to HdfsAuditLogger and other runtime implementation classes as figure \ref{normal_flow} and figure \ref{exc_flow}.

(2) Incomplete control flow
Static analysis is difficult to model the exception handling path and one-step execution flow. Again, for example, in the above function(setPermission) example, AutoLog is unable to generate the above fragment corresponding to the exception thrown, and ignores the "AcessControlException" that may be thrown.
However, the second phase of AnomalyGen identifies the try-catch-finally structure through fine-grained enhanced static analysis, successfully constructs the normal and exception paths and the corresponding log outputs through recursive path merging, validation reasoning (AccessControlException trigger conditions) in the stage three. The graphs of normal and abnoral flow \ref{normal_flow} \ref{exc_flow} clearly shows the control flow, branching conditions, and the corresponding simulated log sequen\-ces.

(3) Lack of dynamic parameter simulation reasoning
Existing tools often use placeholders (e.g., <*>) to represent log parameters, which does not model the actual distribution of runtime parameters. And the lack of parameters may result in anomaly detection tools getting less semantics, which reduces model performance (see what logbase references specifically)
But AnomalyGen accomplishes the data flow constraints during the merge process, and through context and a priori knowledge can simulate the values of the variables based on the semantics. An example is shown below:

\subsection{RQ3: Can AnomalyGen retain contextual semantic information?}  
In contrast to other methods, there is no execution path content corresponding to the log, but we have traced the control flow of the whole execution process as graphs show \ref{normal_flow} \ref{exc_flow}. Through the practice of our staff, we found that the existence of such a control flow diagram will save more than 5 times of the time to find the log and the corresponding control flow than not.


\subsection{RQ4: Can AnomalyGen benefit anomaly detection problems?}
Experimental design:
Three anomaly detection models, including Transformer, CNN, and LSTM, are trained on the baseline dataset with and without AnomalyGen dataset respectively.

Compare the performance of the models on Precision, Recall, and F1-score in the HDFS datasets, and analyze the generalization ability in complex scenarios.

Experimental results:
The models trained on the data augmented by AnomalyGen dataset have an average F1-score improvement of about 1.8\% (e.g., LSTM improved 3.7\% from 0.917 to 0.954). The performances of the dataset enhanced by AnomalyGen are improved when it is trained in all three advanced anomaly detection models, which fully demonstrates that the dataset generated by AnomalyGen is of high quality and beneficial to the development of anomaly detection technology.  

\begin{table*}[!htb]
    \caption{COMPARISON OF ANOMALY DETECTION MODELS OVER BASELINE DATASET AND BASELINE DATASET AUGMENTED WITH DATA GENERATED BY ANOMALYGEN}
    \label{tab:3}
    \centering
        \begin{tabular}{c|ccccccc|ccccccc}
        \hline
        & F1-Score  & Precision  & Recall  \\ 
        & (Without/With AG) & (Without/With AG) & (Wtihout/With AG) \\
        \hline
        LSTM & 0.917/\textbf{0.954} & 0.890/\textbf{0.961} & 0.944/\textbf{0.947} \\ 
        Transformer & 0.852/\textbf{0.867} & 0.865/\textbf{0.964} & 0.839/0.788 \\
        CNN & 0.969/\textbf{0.971} & 0.945/0.926 & 0.992/\textbf{0.996} \\
        \hline
        \end{tabular}
\end{table*}


\section{Threats to Validity}
We will address the following three main validity threats:

\textbf{Dynamic Parameter Resolution Limitations (DPRL).} 
Although AnomalyGen generates dynamic parameters (e.g., variable values, number of cycles) via LLM, the parameter distribution may deviate from the real scenario. For example, LLM may generate data that does not match the display (e.g., a MB size may be set to a value regardless of out-of-memory), resulting in less realistic logs. In addition, static analysis cannot fully parse parameters in reflection calls or dynamic proxies, resulting in some missing log parameters.
Mitigation: In the future, we will use symbolic execution techniques to derive possible ranges of values for parameters in Phase II's Enhanced CFG generation and use them as input constraints for LLM to improve the reliability of parameter prediction.

\textbf{LLM Reasoning Uncertainty.}
AnomalyGen relies on LLM for control-flow reasoning. However, there is a certain amount of randomness and logical errors in its output. For example, LLM may incorrectly complete missing call nodes or generate unreachable paths in conditional branches. In addition, LLM is limited by the number of tokens, which makes it difficult to maintain a large amount of valid information, and to reason about and memorize particularly long sequences.
Mitigation: We consider adding multiple rounds of inference and expert rule filtering to reduce errors in LLM output in future CoT validation in Phase III. In addition, add memory and compression mechanisms to help analyze long sequences.

\textbf{Incomplete Anomaly Annotation Rules.}
Phase IV's anomaly annotation relies on explicit and implicit rules, but some anomaly scenarios (e.g., emerging error codes, custom anomaly types) may not be covered by the rules. In addition, the subjective nature of expert annotation may lead to inconsistent annotations (e.g., differences in judgments about "potential exceptions").
Mitigation: We will regularly update the anomaly rule base to include newly discovered anomaly patterns. During the expert annotation stage, we adopt the majority voting mechanism and continuously monitor the annotation consistency through Krippendorff's α coefficient to ensure the quality of annotation.




\section{Related Work}




\subsection{Log Statement Generation}
Developers write log statements to generate logs and record the execution behavior of the system to assist in debugging and software maintenance \cite{li2020shall}. However, it is a challenge to obtain sufficient, high-quality and representative logs for practical analysis \cite{huo2023autolog}. Automatic log generation frameworks based on technologies such as deep learning \cite{mastropaolo2022using,ding2022logentext,ding2023logentext}, program analysis \cite{huo2023autolog,li2024go}, and large model generation \cite{li2024exploring,xu2024unilog} provide a solution to this challenge.

In terms of deep learning models, Mastropaolo et al. \cite{mastropaolo2022using} proposed LANCE, which used a deep learning model to take Java methods as input and inject complete log statements into them. Ding et al. expanded LoGenText\cite{ding2022logentext} into LoGenText-PLus \cite{ding2023logentext} by incorporating syntactic templates of log text.

In terms of program analysis, Huo et al. \cite{huo2023autolog} proposed the first automatic log generation method for anomaly detection, AutoLog, which uses program analysis to generate runtime log sequences without actually running the system. Experiments have proven that AUTOLOG enables log-based anomaly detectors to achieve better performance than existing log datasets. Li et al. \cite{li2024go} proposed SCLogger, which is the first context-aware log statement generation method with inter-method static context.

Inspired by the great success of LLMs in natural language understanding and code intelligence tasks, some scholars have begun to explore the advantages of LLMs in log generation. Li et al. \cite{li2024exploring} were the first to conduct an exploration and empirically analyze the performance of LLMs in generating log statements. Xu et al. \cite{xu2024unilog} proposed the UniLog framework, which is the first universal framework that utilizes the context learning ability of LLMs to achieve automatic, end-to-end logging. 





\subsection{Log Anomaly Detection}
Traditional log anomaly detection research has explored log anomaly detection from perspectives such as graph structures \cite{li2024graph} and semantic analysis \cite{huo2023evlog, wang2023deepuserlog, shavit2024semantilog}. Li et al. \cite{li2024graph} converted event logs into attributed, directed, and weighted graphs, and then constructed a graph neural network model to detect graph - level anomalies in a set of attributed, directed, and weighted graphs. Huo et al. \cite{huo2023evlog} proposed the first study on locating abnormal logs during software evolution. Wang et al. \cite{wang2023deepuserlog} detected anomalies in user logs containing a large number of key-value pairs, extracting semantic features of the content after removing the values in key-value pairs. Shavit et al. \cite{shavit2024semantilog} regarded log - based anomaly detection as a semantic similarity problem and detected anomalies through the cosine similarity between encodings.

Some scholars are committed to exploring the possibilities of deep learning models in log anomaly detection \cite{guo2024logformer, li2024multi}. Guo et al. \cite{guo2024logformer} proposed a framework called LogFormer to improve the generalization ability of log anomaly detection tasks across different domains. Li et al. \cite{li2024multi} designed a feature extraction module for multi - source log data, effectively capturing the features of log data and key performance indicator data.

Against the backdrop of the popularity of deep learning and large models, some scholars have attempted to improve traditional methods \cite{yu2024deep, yang2024try}, demonstrating the possibility that traditional methods can outperform deep - learning methods. Yu et al. \cite{yu2024deep} evaluated the performance of basic algorithms and deep - learning methods on five public log anomaly detection datasets. The results showed that simple algorithms outperformed deep - learning methods in terms of both time efficiency and accuracy. Yang et al. \cite{yang2024try} conducted the first empirical study by incorporating the lightweight semantics - based log representation method, SemPCA, into traditional PCA technology . Experiments have shown that SemPCA achieved better performance than the best deep learning techniques.


\subsection{Log Analysis Based on Large Language Models}
As semi-structured text, logs are rich in semantic information. With the success of pre-trained language models in the field of natural language processing, many studies have utilized these models for log analysis. Ma et al. \cite{ma2024knowlog} proposed the knowledge-enhanced pre-trained language model KnowLog for log understanding, which demonstrated significant advantages in transfer learning and low-resource scenarios. Liu et al. \cite{liu2024interpretable,liu2024logprompt} proposed a novel interpretable log analysis method for online scenarios, LogPrompt, which utilizes large language models and advanced prompt techniques to achieve performance improvement in zero-shot scenarios. 

\section{Conclusion}

With the advantages of comprehensiveness, realism, and contextual integrity, AnomalyGen-generated logs significantly improve the performance of anomaly detection models by effectively generating more comprehensive and realistic log data than any existing Java log dataset. It can be used as a benchmark data generator, providing a solid data foundation for the development of anomaly detection techniques.


\newpage
\bibliographystyle{ACM-Reference-Format}
\bibliography{software.bib}










\end{document}